\documentclass[10pt,letterpaper]{article}
\usepackage[OE]{express}

\usepackage{ulem}
\usepackage{siunitx}
\DeclareSIUnit\decibelc{dBc}
\usepackage{amsmath,subfigure}
\allowdisplaybreaks
\usepackage{multirow}

\begin{document}

\title{Coherent multi-mode dynamics in a Quantum Cascade Laser: Amplitude- and Frequency-modulated Optical Frequency Combs}

\author{Carlo Silvestri$^{1,*}$, Lorenzo Luigi Columbo$^{1}$, Massimo Brambilla$^{2}$, Mariangela Gioannini$^{1}$}





\address{
$^1$ Dipartimento di Elettronica e Telecomunicazioni, Politecnico di Torino, Corso Duca degli Abruzzi 24, Torino, IT-10129, Italy\\
$^2$ Dipartimento Interateneo di Fisica, Politecnico ed Universit\`a\ degli Studi, Via Amendola 173, Bari, IT-70126, Italy}
\begin{center}
\email{*carlo.silvestri@polito.it}
\end{center}

\begin{abstract}
We cast a theoretical model based on Effective Semiconductor
Maxwell-Bloch Equations and study the dynamics of a multi-mode mid-Infrared Quantum Cascade Laser in Fabry Perot with the aim to investigate the spontaneous generation of optical frequency combs. This model encompasses the key features of a semiconductor active medium such as asymmetric, frequency-dependent gain and refractive index as well as the  phase-amplitude coupling of the field dynamics provided by the linewidth enhancement factor. Our numerical simulations are in excellent agreement with recent experimental results, showing broad ranges of comb formation in locked regimes, separated by chaotic dynamics when the field modes unlock. In the former case, we identify self-confined structures travelling along the cavity, while the instantaneous frequency is characterized by a linear chirp behaviour. In such regimes we show that OFC are characterized by concomitant and relevant amplitude and frequency modulation.
\end{abstract}


\bibliographystyle{osajnl}

\section{Introduction}

Quantum Cascade Laser (QCLs) have attracted a remarkable interest as THz and Mid-IR sources  capable of self-starting Optical Frequency Combs (OFCs) under DC current injection \cite{faist,faistnat,Scalari1,Burghoff}. OFCs are generally meant as lasers emitting, under particular  bias conditions, a set of equally spaced optical lines with low phase and amplitude noise \cite{udem}. These optical sources are  appealing for a wealth
of applications in the Mid-IR and THz range, encompassing high precision molecular spectroscopy, broad band free space optical communication and hyperspectral imaging \cite{revcombs, bartalini}.\\
From an experimental point of view, the OFC regime has been mainly characterised through the intermode beatnote (BN) spectroscopy and associated with a narrow BN linewidth (typically less than 100KHz). By sweeping the bias current, it was found bias current ranges  of irregular dynamics (phase unlocked optical and wide BN linewidth) alternated with current ranges of  OFC operation (phase locked lines and narrow BN linewidth)  \cite{revcombs, barbieri,Wienold}.
Figures of merit of the OFC are typically the number of locked modes in the $-40dB$ (or  $-20dB$) spectral bandwidth and the OFC dynamic range, intended as the range of bias current where OFC  emission occurs.  In this regard, THz QCLs emitting at $3.1THz$ can provide up to few tens of modes in the  $-40dB$ spectral bandwidth of about $1.1THz$; whereas Mid-IR QCL can give self-locked optical lines in a bandwith of about  $3$THz centered at $36.5$THz \cite{scalari2018,Otticafaist}.   In absence of any dispersion compensation \cite{villares3,Mezzapesa} or microwave modulation \cite{barbierimicro}, stable OFC regimes have been found in ranges of current of about one hundred milliamperes starting from about two times the lasing threshold \cite{scalari2018,Otticafaist}.
Only recently the temporal dynamics of the optical field  became accessible through the Shifted Wave Interference Fourier Transform Spectroscopy (SWIFTS) technique \cite{Burghoff} that allows to retrieve the amplitude and phase of the optical field from experimental data \cite{Otticafaist, Schwarz}. This additional information revealed the true nature of the self-generated OFC in QCLs: it occurs not only in presence of a Frequency Modulated (FM) laser emission, but its formation also implies   a significant (or even dominant) Amplitude Modulation (AM), appearing as intra-cavity optical pulses which propagate on a quasi-homogeneous background field \cite{Otticafaist, Burghoff2}. In addition, the inspection of the temporal evolution of the field phase and the consequent instantaneous frequency, demonstrates the existence of linear frequency chirp with a frequency jump at the time instants where the field amplitude is modulated by the pulse \cite{Otticafaist, Burghoff2,Schwarz}.
Well before experimental SWIFTS measurements, theoretical predictions of such pulses having a solitary wave character was provided in \cite{columbo2018}.\\
Although several theoretical efforts  have been made in order to provide a physical understanding of the fascinating phenomenon of self-starting OFCs in QCLs, to the best of our knowledge there is still a lack of models able to reproduce the experimentally measured coexistence of optical pulses and linear frequency chirp, and also the alternance between locked and unlocked regimes. We believe that such tools would be promising to  predict  possible strategies to extend the OFC dynamic range by employing externally controllable signals, by optimizing the device gain material or the laser cavity design.
The approaches proposed so far are based on the classical set of Maxwell-Bloch equations valid only for an ideal two o three levels atom-like material system \cite{villares, Tzenov, capasso0, Kurgin, Boiko}. This model, while grasping some basic features of the laser physics,  fails in correctly describing the  phase-amplitude coupling (quantified by the linewidth enhancement factor, LEF ) peculiar of semiconductor lasers.
In absence of the phase-amplitude coupling,  the relevant mechanism in determining the multimode instability threshold and influencing the possibility to observe OFCs was only ascribed to the Spatial Hole Burning (SHB) \cite{Gordon, Boiko, LugiatoPhysScr} consisting in a  carrier grating excited by the interfering counter propagating field of the Fabry-Perrot (FP) laser cavity.
More recently,   a non-null LEF  and an inhomogeneous gain broadening have been "ad hoc"  included in  \cite{Burghoff2, Schwarz}; new kind of CW instability and  multi-mode dynamics have been found with a better match with some of the experiments reported for e.g. in \cite{Otticafaist}   \\

In \cite{columbo2018} we adopted a model consisting on a set of Effective Semiconductor Maxwell-Bloch Equations (ESMBEs) \cite{Ning} to study THz QCLs and we demonstrated it could well reproduce the experimental observation of self-generated OFCs alternated with ranges of irregular multi-mode regimes \cite{barbieri}. The  ESMBEs was based on a non-linear optical susceptibility model that describes radiation-matter interaction by fitting microscopic calculated and/or experimentally measured optical gain spectra and refractive index dispersion.  This allowed us to point out the role played by the LEF in  reproducing  and explaining:
\begin{itemize}
  \item  the instability of the CW lasing even close to lasing threshold, whereas it was originally predicted to occur above about ten times the threshold current due to the Risken-Nummedal-Graham-Haken (RNGH) instability
  \item the multi-mode dynamics, due to the onset of solitary pulses travelling in the resonator, and narrow  BN spectra at the round trip frequency (or at the double of the BN explaining the disappearance of the BN in some  experiments)
  \item the self-generation of OFCs not only close to the threshold current but also in current ranges beyond regions of  irregular unlocked dynamics
\end{itemize}
Since  the model in \cite{columbo2018}  was rigorously valid only in a unidirectional ring configuration and with no coupling to the output waveguide, the aim of this work is extending that approach to the case of a more realistic FP laser by including in the model the formation of the standing wave field pattern along the laser cavity through the additional presence of the SHB caused by the carrier grating \cite{Gordon, bardella2017}. \\
We apply the model to study Mid-IR FP QCL and we show that simulation results are in  very good agreement with the experimental evidences reported  in \cite{Otticafaist}. Specifically, we can exhaustively describe the phenomenon of self-starting OFCs and the coexistence of the AM and FM regime characterized by weak pulses with the cavity round trip repetition rate and FM linear chirp of the output optical field. Thanks to a campaign of simulations exploring the parameter space of LEF, carrier lifetime and optical gain bandwidth, we can show the impact of these parameters on the extension of the current range of self-OFC generation and on the number of locked optical modes. \\
In Section 2 we derive the model for a FP QCL, remarking the role of the carrier grating in the medium dynamics and linking the model parameters to physical quantities, relevant for comparison with experiments.
In Section 3 we start by illustrating results from our model, relative to a realistic case. We show that laser field, upon current ramping, can exhibit locking regimes, multiple locking windows, window of chaotic dynamics, AM and FM dynamics associated thereof as well as instantaneous frequency chirping (FCh), with close similarity to experimental evidence. Some quantifiers are introduced, to characterize both  OFC formation (beyond standard reference to BN linewidth)  and FCh regimes.
Following this reference case, in paragraph 3.1 we illustrate the role of critical medium parameters, mainly the LEF and the gain bandwidth, in ruling the OFC regime extension, the spectral character of the OFC and we compare our evidence to laser dynamics in absence of SHB and approaching the two-level case.
Finally in paragraph 3.2 the role of carrier rates is considered, confirming that 'slower' carriers give rise to longer pulses and eventually lead to the loss of the pulsed regime. Conversely, faster carriers give rise to shorter pulses and we can show that, interestingly,  the corresponding simulations also reveal the formation of OFC encompassing larger number of modes, occurring for larger gain linewidth and in ampler current ranges, thus in very good agreement with the results recently reported for e.g. in \cite{Otticafaist}.
Sec.4 draws conclusions and prospects future   developments.

\section{The model. Effective Semiconductor Maxell-Bloch Equations for a Fabry-Perot multimode QCL}


Our model encompasses the semiconductor susceptibility, typical of a QCL, originally developed in \cite{columbo2018} for an unidirectional resonator, combined with the multiple scale approach adopted for Quantum Dot (QD) lasers in \cite{bardella2017} to account for carriers grating due to standing wave pattern and responsible for Spatial hole burning, with the goal to properly describe a bidirectional Fabry-Perot (FP) resonator (see also chapter 14 in \cite{NOS}). \\
We consider a FP cavity a few millimeters long and start by treating the spatio-temporal evolution of the electric field. We start from the d'Alembert equation
\begin{equation}
\frac{\partial^2 E}{\partial z^2}- \frac{1}{v^2}\frac{\partial^2 E}{\partial t^2}=\frac{1}{\epsilon_0 c^2}\frac{\partial^2 P}{\partial t^2} \label{dal}
\end{equation}
where $E$ is the electric field, $P$ is the polarization and $v$ is the radiation group velocity. We then assume that the electric field and the polarization can be expressed as
\begin{equation}
E(z,t)=\frac{1}{2}[E(z,t)^+\exp(-ik_0z+i\omega_0 t)+E(z,t)^-\exp(+ik_0z+i\omega_0 t)+c.c.] \label{el1}\\
\end{equation}
\begin{equation}
P(z,t)=\frac{1}{2}[P_0(z,t)\exp(+i\omega_0 t)+c.c.] \label{pol1}
\end{equation}
where $E^+(z,t)$, $E^-(z,t)$ are respectively the slowly varying envelopes for the forward field and backward field inside the resonator and $P_0(z,t)$ is the polarization envelope, assumed to vary slowly only in time for reasons that will be clarified in the following passages,$\omega_0$ and $k_0$ are respectively the reference frequency (cold cavity mode closest to the gain peak) and its wavenumber.\\
Inserting Eqs. (\ref{el1}) and (\ref{pol1}) in Eq. (\ref{dal}) and applying the slowly varying envelope approximation (SVEA) we obtain the following equation:
\begin{equation}
\left[ \frac{\partial E^+}{\partial z}+ \frac{1}{v}\frac{\partial E^+}{\partial t}\right]\exp\left(-ik_0z \right)+\left[- \frac{\partial E^-}{\partial z}+ \frac{1}{v}\frac{\partial E^-}{\partial t}\right]\exp\left(+ik_0z \right) = g P_0 \label{el2}
\end{equation}
where g is a complex coefficient given by:
\begin{equation}
g=\frac{-i\omega_0N_p\Gamma_c}{2\epsilon_0nc} \, , \label{g}
\end{equation}
$N_p$ is the number of stages in the cascading scheme, $\Gamma_c$ is the optical confinement factor (that takes into account the overlap between the optical mode and the active region) and $n$ is the effective background refractive index of the medium.\\
The field dynamics is coupled to the active medium's one. Starting from the carrier density, we assume that in each transition, within the cascaded  superlattice, the ground state is always empty, because of the fast depopulation due to the LO phonon-electron scattering processes. Therefore in our model only the carrier density of the upper laser level $N(z,t)$ appears as a dynamical variable.  The evolution equation is retrieved from the Bloch two-level approach \cite{NOS} in the rotating wave approximation. We consider a pumping current $I$, the carrier nonradiative decay time $\tau_e$, and take into account the forward and backward field envelopes, as it is required for FP cavities. We obtain:
\begin{eqnarray}
\frac{\partial N}{\partial t}&=&\frac{I}{eV}-\frac{N}{\tau_e}-\frac{i}{4\hbar}\left[ \left(E^+exp(-i k_0 z)+E^-exp(+ik_0z)\right)P_0^*\nonumber \right. \\
&-&\left. \left(E^{+*}exp(+ik_0z)+E^{-*}exp(-ik_0z)\right)P_0 \right] \label{Ntot}
\end{eqnarray}
where $V$ is the medium volume and $e$ is the electron  charge.\\

The equation for the polarization dynamics is derived following the approach described in detail in Sec.2 of \cite{columbo2018}. We start by introducing a phenomenological optical susceptibility $\chi(\omega,N)$ that allows to describe spectrally asymmetric curves for gain and  dispersion, generally dependent on the carrier density; it has the form \footnote{We use a different sign convention respect to \cite{columbo2018}, due to the assumptions for the expression of the complex electric field and polarization (Eqs. (\ref{el1})-(\ref{pol1})).}:
\begin{equation}
\chi(\omega, N)=\frac{f_0N\left(1+i\alpha\right)\left(i-\alpha\right)}{\left(1+i\alpha\right)+i\omega \tau_{d}}\label{susc}
\end{equation}
In Eq. (\ref{susc}) we have assumed, for simplicity, that the gain maximum coincides with the reference cavity frequency $\omega_0$ \footnote{The FSR is large enough so that a moderate frequency shift of the gain peak is of little relevance to the laser dynamics.}.
Eq. (\ref{susc}) is associated in the time domain to the following polarization equation where the peculiar feature of the FP resonator is made evident by the dependency from the counterpropagating field envelopes:
\begin{eqnarray}
\frac{\partial P_0}{\partial t}&=&\frac{1}{\tau_d}(1+i\alpha)\left[-P_0+if_0\epsilon_0\epsilon_b\left(1+i\alpha\right)N\left(E^+\exp\left(-ik_0z\right)+E^-\exp\left(+ik_0z\right)\right) \right] \label{Ptot}
\end{eqnarray}
where $\alpha$ is the LEF and $\frac{1}{\tau_d}$ is the effective polarization decay rate \footnote{Note that the effective polarization decay time corresponds to $\frac{\Gamma}{\tau_d}$ in Eq.(13) of \cite{columbo2018}.}. For further convenience we introduce $\delta_{hom}=\frac{1}{\pi \tau_{d}}$, which is a measure of the FWHM of the gain spectrum in the limit  $\alpha << 1$ where the susceptibility $\chi(\omega,N)$ becomes that of homogeneous broadened two-level system gain \cite{columbo2018}.\\

At this point our equations include field-carrier interactions at all spatial orders (measured in multiples of $\lambda$), but in order to retain physical insight and numerical viability, a relevant simplification can be introduced by exploiting a multiple scale approach \cite{Homar, Serrano, Stenholm}. Specifically, we expand in Fourier series the spatial variation at the wavelength scale of $P$ and $N$ \cite{Stenholm}:
\begin{eqnarray}
P_0=exp(-ik_0z)\sum_{n=0}^{\infty}{P_n^+exp\left(-2nik_0z\right)}+exp(+ik_0z)\sum_{n=0}^{\infty}{P_n^-exp\left(+2nik_0z\right)}\label{Pfour}\\
N=N_0+\sum_{n=1}^{\infty}{N_n^+exp\left(-2nik_0z\right)}+\sum_{n=1}^{\infty}{E_n^-exp\left(+2nik_0z\right)}\label{nfour}
\end{eqnarray}
Inserting Eqs. (\ref{Pfour}) and (\ref{nfour}) into Eqs. (\ref{el2}), (\ref{Ntot}) and (\ref{Ptot}) and neglecting the terms with spatial frequency higher than $2k_0$, we get the final set of Effective Semiconductor Maxwell-Bloch Equations (ESMBEs) for QCL in FP configuration in the form:
\begin{eqnarray}
\frac{\partial E^+}{\partial z}+ \frac{1}{v}\frac{\partial E^+}{\partial t} &=& -\frac{\alpha_L}{2}E^++g P_0^+ \label{el+}\\
-\frac{\partial E^-}{\partial z}+ \frac{1}{v}\frac{\partial E^-}{\partial t} &=& -\frac{\alpha_L}{2}E^-+g P_0^- \label{el-}\\
\frac{\partial P_0^+}{\partial t}&=&\frac{(1+i\alpha)}{\tau_d}\left[-P_0^++if_0\epsilon_0\epsilon_b\left(1+i\alpha\right)\left(N_0E^{+}+N_1^+E^-\right) \right] \label{P+}\\
\frac{\partial P_0^-}{\partial t}&=&\frac{(1+i\alpha)}{\tau_d}\left[-P_0^-+if_0\epsilon_0\epsilon_b\left(1+i\alpha\right)\left(N_0E^{-}+N_1^-E^+\right) \right] \label{P-}\\
\frac{\partial N_0}{\partial t}&=&\frac{I}{eV}-\frac{N_0}{\tau_e}+\frac{i}{4\hbar}\left[E^{+*}P_0^++E^{-*}P_0^--E^{+}P_0^{+*}-E^{-}P_0^{-*}\right] \label{N0}\\
\frac{\partial N_1^+}{\partial t}&=&-\frac{N_1^+}{\tau_e}+\frac{i}{4\hbar}\left[E^{-*}P_0^+-E^{+}P_0^{-*}\right] \label{N1}
\end{eqnarray}
Finally, the model equations must be completed by the boundary conditions which read:\\
\begin{eqnarray}
E^-(L,t)&=&\sqrt{R}E^+(L,t)\label{bc1}\\
E^+(0,t)&=&\sqrt{R}E^-(0,t)\label{bc2}
\end{eqnarray}
where $R$ is the reflectivity of each mirror of the FP cavity.

\section{The numerical simulations: Self-generated frequency and amplitude modulated OFCs}
In this section we present the results obtained by numerical integration of the ESMBEs (\ref{el+}-\ref{N1}) with the boundary conditions (\ref{bc1})-(\ref{bc2}) for typical Mid-IR QCL parameters reported in Table 1 and adopted from literature \cite{faist,Otticafaist}. The numerical code is based on a TDTW algorithm, which exploits a finite differences scheme, discretizing both in time and space \cite{bardella2017}.
\begin{figure}
\centering
\includegraphics[width=0.80\textwidth]{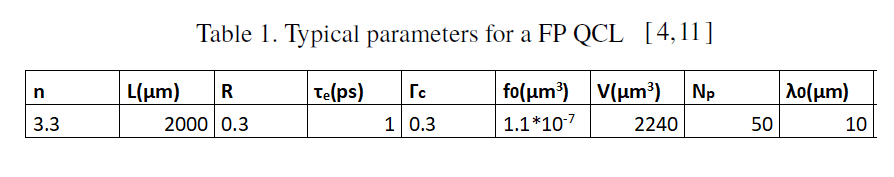}
\label{Domain1}
\end{figure}
Our first aim is the reproduction of OFC regimes with characteristics similar to those experimentally observed \cite{vitiello,barbieri,Otticafaist,Scalari2016, scalari2017}, namely: a combination of FM and AM OFCs occurring close to the lasing threshold and in a significant bias current range, followed by a current range of unlocking with irregular dynamics and, possibly, occurring again in a second window for larger bias currents, a feature that is commonly observed in experiments, but that, to the best of our knowledge, was never found theoretically.\\
In such perspective, we first present the typical results adopting  the realistic values of $\alpha=0.4$ and $\delta_{hom}=0.48$THz. The corresponding light-current plot is reported in Fig. \ref{LIcurve}.
\begin{figure}[!h]
\centering
\includegraphics[width=0.55\textwidth]{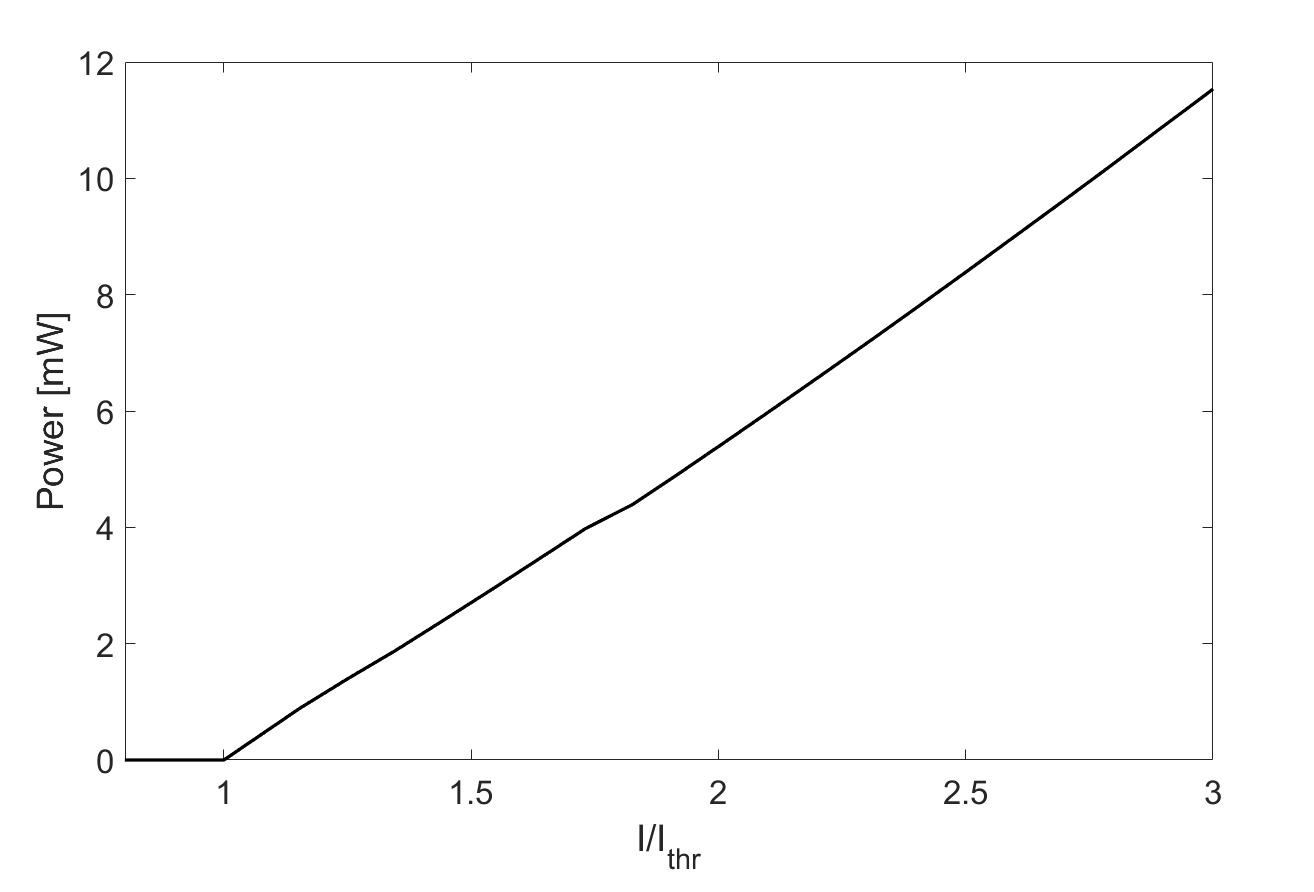}
\caption{Power as a function of the ratio $I/I_{thr}$ for $\alpha=0.4$, $\delta_{hom}=0.48$THz. In this case $I_{thr}=260mA$. Power is the time average over a simulation time window of about $500ns$, after a stable regime is attained. the other parameters are those reported in Table 1.}
\label{LIcurve}
\end{figure}
Further on, we will present the results of a massive campaign of simulations showing  a broad  zoology of dynamical regimes and the impact of the LEF, optical gain bandwidth and carrier lifetime on the figure of merit of the self-generated OFC.  \\
We first focus on the identification of OFC regimes by sweeping the bias current $I$. In our simulations, the emergence of a OFCs regime can be characterized, as typically done in experiments, by a narrow BN linewidth at Radio Frequency (RF). However a better assessment can be achieved by estimating some additional phase and amplitude noise quantifiers that we have recently introduced for the numerical characterization of OFCs  in  QD lasers \cite{bardella2017}. To calculate them, the spectrum of the optical field at z=L (exit facet of the simulated device) is filtered so as to retain only the modes within a 10dB power ratio to the spectral maximum. We then consider the temporal evolution of each filtered optical line of the spectrum: the modal amplitudes $P_q(t), q=1,...,N_{10}$ and the temporal phase difference between one mode and the adjacent one $\Delta\Phi_q(t), q=1,...,N_{10}$, where $N_{10}$ is the number of optical lines in the $-10dB$ spectral bandwidth \cite{bardella2017}.
Given the amplitude and phase dynamics of each optical line, we  calculate the quantities:
\begin{equation}
M_{\sigma_{P}}=\frac{1}{N_{10}}\sum_{q=1}^{N_{10}}{\sigma_{P_q}}\quad , \quad M_{\Delta\Phi}=\frac{1}{N_{10}}\sum_{q=1}^{N_{10}}{\sigma_{\Delta\Phi_q}}\label{MsigmaMPhi}
\end{equation}
where:
\begin{equation}
\mu_{P_q}=<P_q(t)> \quad , \quad
\mu_{\Delta\Phi_q}=<\Delta\Phi_q(t)>
\end{equation}
\begin{equation}
\sigma_{P_q}=\sqrt{<\left(P_q(t)-\mu_{P_q}\right)^2>} \quad , \quad
\sigma_{\Delta\Phi_q}=\sqrt{<\left(\Delta\Phi_q(t)-\mu_{\Delta\Phi_q}\right)^2>}
\end{equation}
and the symbol $< \, >$ indicates the temporal average.\\
 The indicators defined by Eq.\ref{MsigmaMPhi} measure the average fluctuations of the power and phase of the selected optical lines. An ideal OFC should have no intensity noise fluctuation of the power of each line (ie: low RIN per line) and zero differential  phase noise such  that both indicators should be zero. In our simulations we observe residual fluctuations, so that we will define in the following an OFC regime when the indicators are $M_{\sigma_{P}}<10^{-2}mW$ and $M_{\Delta\Phi}<2\cdot10^{-2} rad$.\\
 
\begin{figure}[!h]
\centering
\includegraphics[width=0.95\textwidth]{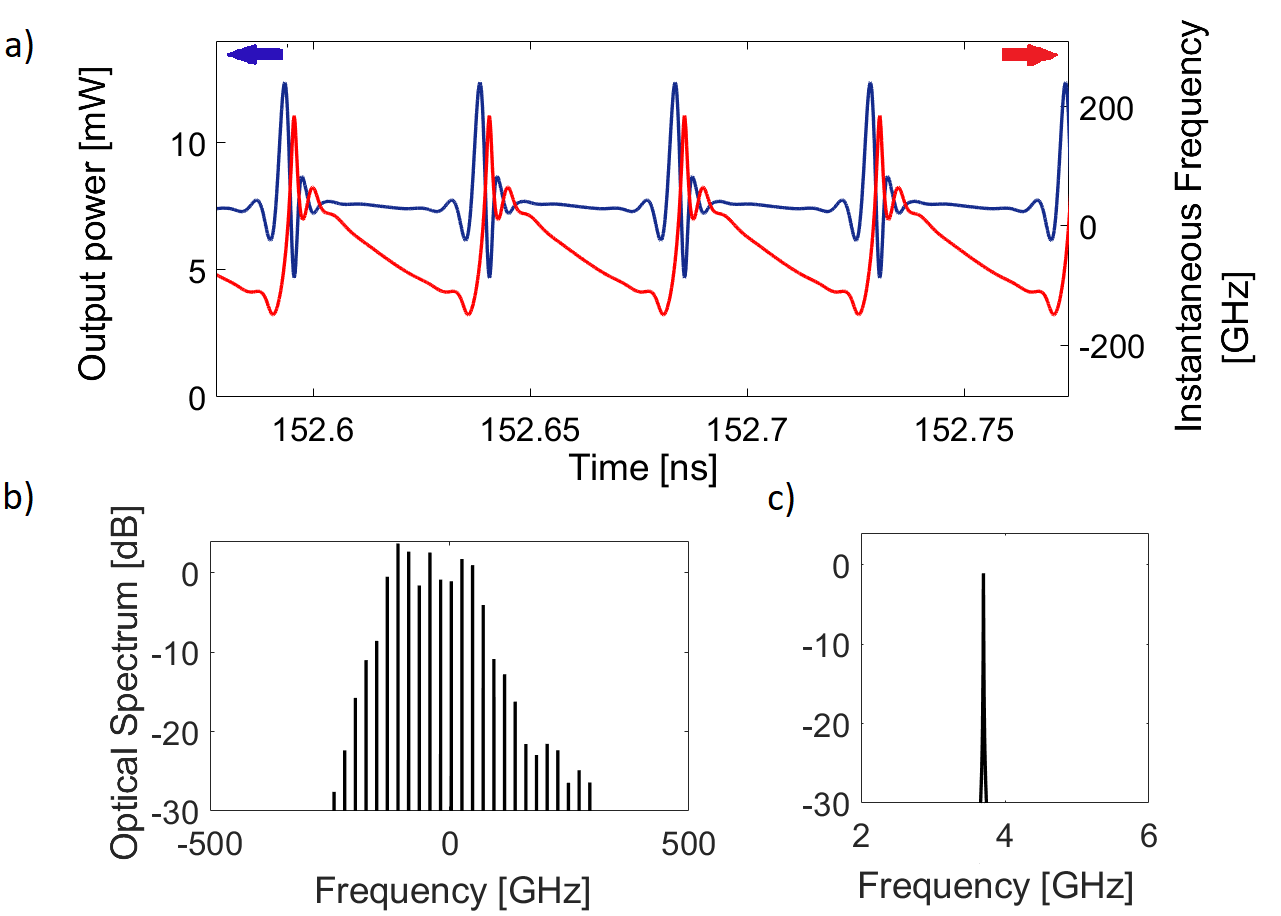}
\caption{OFCs emission for $I/I_{thr}=2.31$. Here $\alpha=0.4$, $\delta_{hom}=0.48THz$, other parameters as in Table 1. Temporal evolution of laser power (blue curve) and instantaneous frequency (red curve). A propagating pulse at the round trip frequency sits on an almost constant background associated with a linear frequency chirp. (b) Optical spectrum of the emitted radiation showing $10$ modes in the $-10dB$ spectral bandwidth. c) Zoom around one peak of the optical spectrum.}
\label{regular}
\end{figure}

 An example of dynamical behaviour corresponding to the self-starting OFC is shown in Fig. \ref{regular}, for $I/I_{thr}=2.31$, where $I_{thr}$ is the threshold current of the laser. The propagation of confined field structures 
 sitting on an almost constant background in the intensity trace (Fig. \ref{regular}.a)
 appears intrinsically paired with an instantaneous frequency chirp in the time range where the intensity is almost constant, followed by discontinuous jumps when the field structure occurs
 (Fig. \ref{regular}.a); note the remarkable similarity with the experimental evidences in Fig. 2.b of \cite{Otticafaist}. This evidence suggests that OFC is a locking phenomenon where concomitant AM and FM is a commonplace (see also fig.8). \\
 
\begin{figure}[h!tb]
\centering
\includegraphics[width=0.95\textwidth]{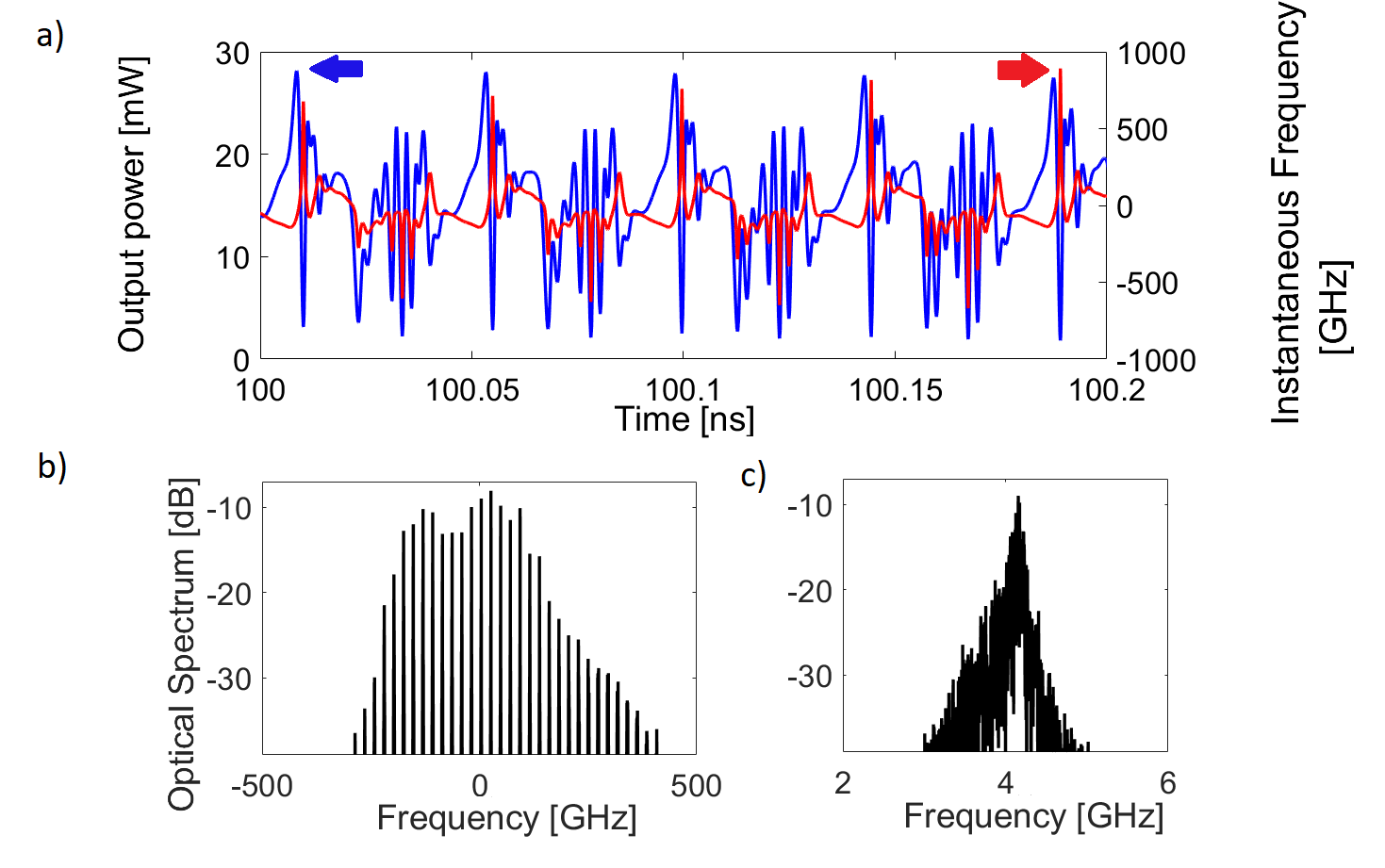}
\caption{Chaotic behaviour for $I/I_{thr}=3.46$. Other parameters are as in Fig. \ref{regular}. (a) Temporal evolution of laser power (blue curve) and instantaneous frequency (red curve). Irregular oscillations can be easily detected. (b) Optical spectrum of the emitted radiation. c) Zoom around one peak of the optical spectrum.}
\label{irregular}
\end{figure}

 Additionally, we observe $10$ locked lines in the $-10dB$ spectral bandwidth of 0.2 THZ (Fig. \ref{regular}.b); each line has a very narrow  linewidth as shown by the zoom around one line in Fig. \ref{regular}.c\\
 When the laser unlocks, an irregular dynamics is observed as for example at $I/I_{thr}=3.46$. The field intensity and its instantaneous frequency versus time are shown in Fig. \ref{irregular}a; whereas the whole optical spectrum of Fig. \ref{irregular}b is apparently not too different from Fig. \ref{regular}.b, we note that each line is significantly enlarged with several side bands close to the main peak (Fig. \ref{irregular}.c)).

 Fig. \ref{regular} indicates, in excellent agreement with experimental evidence, that the OFC regime with a broad and flat optical spectrum  \textit{is characterized by an almost linear frequency chirp}. To quantify the linearity of the chirp  at different bias currents and/or for different sets of parameters, we introduce here an  indicator of  chirp linearity, based on the comparison of the simulated instantaneous frequency with  a perfect frequency sawtooth\cite{Arfken}.\\
Since the moduli of two adjacent  Fourier coefficients ($c_{n,st}$ and $c_{n+1,st}$) of the Fourier series of an ideal sawtooth stay in the ratio $\frac{|c_{n+1,st}|}{|c_{n,st}|}=\frac{n}{n+1}$; we calculate the Fourier transform of the  instantaneous frequency signal, we define  $c_n$ the peak of each n-th component of the spectrum and the ratio $R_{n}=\frac{|c_{n+1}|}{|c_{n}|}$. The relative error ($\epsilon_{n}$) between $\frac{n}{n+1}$ and $R_{n}$ and its average over $N_c$ components ($\epsilon_{c}$) are defined respectively as:
\begin{equation*}
  \epsilon_{n}  =  \left|\frac{R_{n}-n/(n+1)}{n/(n+1)}\right| ;
  \epsilon_{c}  = \frac{1}{N_c}\sum_{1}^{N_c}  \epsilon_{n}
\end{equation*}

The indicator  $\epsilon_c$ is therefore a relative error aimed at quantifying the discrepancy between the QCL instantaneous frequency signal and an ideal sawtooth. We assume that a regime can be reasonably defined as 'linearly chirped' when $\epsilon_c<10^{-1}$. \\
 \begin{figure}[!h]
\centering
\includegraphics[width=0.90\textwidth]{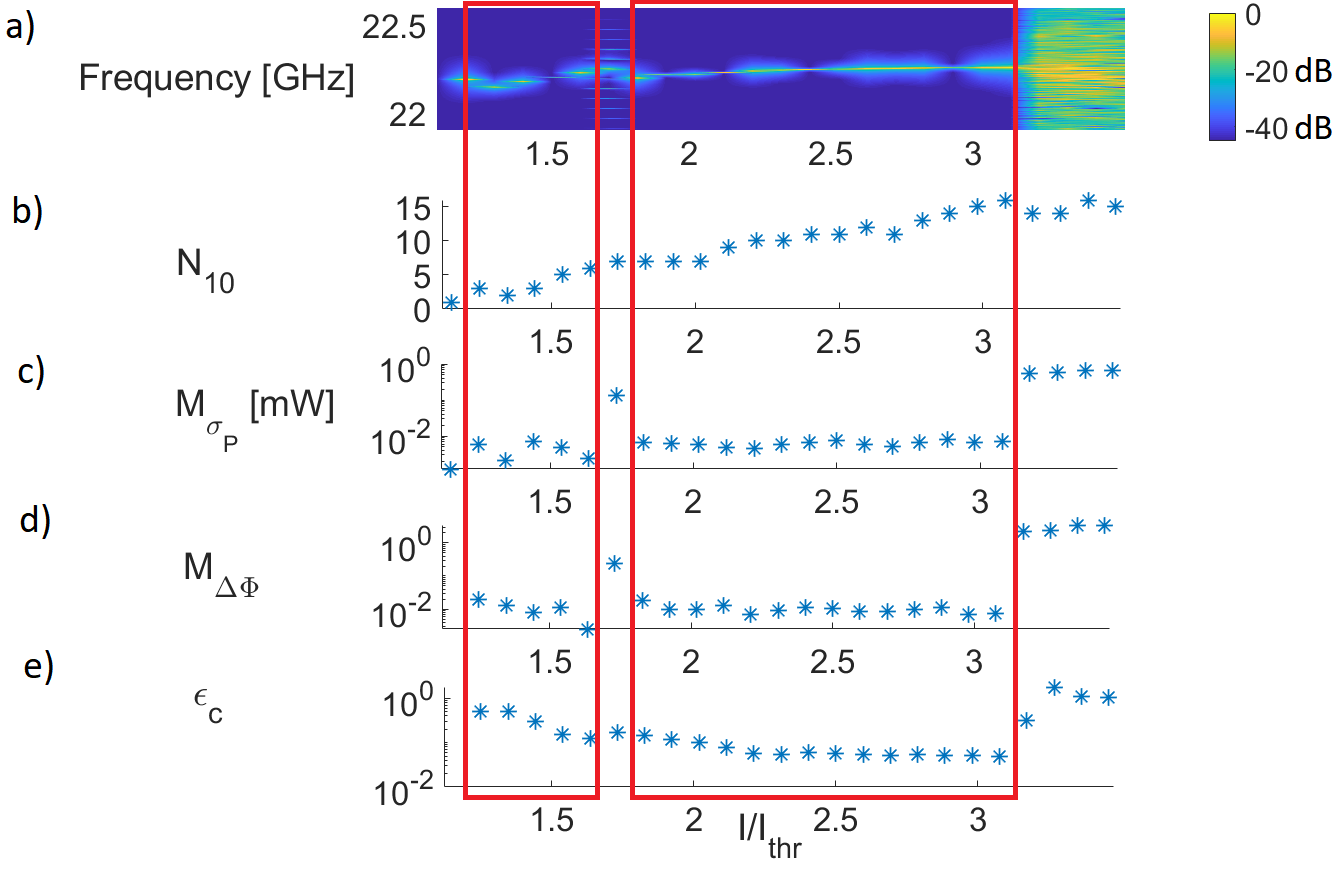}
\caption{Results of simulations for a current scan from QCL threshold $I_{th}$ to $3.5I_{th}$ for $\alpha=0.4$, $\delta_{hom}=0.48THz$. The other parameters are as in Table 1. a) First BN in the RF spectrum (color scale normalized to the maximum for each current value; log scale); b) number of modes in a -10dB spectral bandwidth;
(c) amplitude and (d) phase noise quantifiers for the $N_{10}$ modes, as introduced in \cite{bardella2017}; (e) chirp quantifier for the first $N_{c}=5$ Fourier coefficients of the instantaneous frequency signal. Two regions of OFCs operation highlighted with a red rectangular box can be identified.}
\label{firstscan}
\end{figure}

 As Fig. \ref{firstscan}.a shows, our QCL starts off with a CW emission at threshold $(I_{thr}=260mA)$, which is soon destabilized towards a multimode dynamics associated with the appearance of a BN at $I/I_{thr}$ between $1.25$ and $1.64$.\\
 In this current range we see an OFC regime characterized by a gradual increase of $N_{10}$, low intensity and phase noise (since  $M_{\sigma_{P}}<10^{-2} mW$ and $M_{\Delta \Phi}<2\cdot10^{-2} rad$) and rather large linear chirp indicator ($\epsilon_c>10^{-1}$). We also report a BN shift of $0.03GHz$ around $I/I_{thr}=1.34$, which is in agreement, in terms of order of magnitude, with recent experimental results \cite{MezzapesaOE}.\\
 Around $I/I_{thr}=1.73$ the OFC regime is lost; we observe the onset of several lines around the BN causing an important broadening of the BN linewidth. This broadening is a finger print of an unlocked regime  characterized by an amplitude modulation with a period equal to the inverse of the separation between the BN and adjacent side bands. The corresponding phase and intensity noise indicators increase of nearly two order of magnitude.  This regime ceases just before $I/I_{thr}=1.83$ where a new OFCs regime appears, \textit{thus reproducing the locked/unlocked state alternance found in some experiments} \cite{barbieri}. This regime is even more sizeably extending up to $I/I_{thr}=3.08$, after which chaotic emission sets in. Comparing in  the Fig.\ref{firstscan}.e the linear chirp indicator of  the first and the second locking window, we see that for all current $I/I_{thr}<2$  the value of $\epsilon_{c}$ is higher than $10^{-1}$. In this region $N_{10}$ is less than 9.
In the second locking region for $I/I_{thr}>2$ we have linear chirp with $N_{10}>10$ and an increase of the number of locked modes is accompanied by a further reduction of the linear chirp indicator.
The observed correlation between the reduction of $\epsilon_{c}$ and an increasing number of locked modes suggests that linear chirp is a complex cooperative phenomenon involving a highly multimode dynamics (note that in calculating our $\epsilon_{c}$ we choose $N_{c}$=5).\\
As proposed in \cite{revcombs} the spontaneous formation OFCs is due to efficient Four Wave Mixing (FWM) that for sufficiently high interactivity field intensity (or bias current) acts as a self-injection locking mechanism in compensating the cavity mode dispersion and fixing their relative phase differences.\\

\subsection{OFCs properties: the role of LEF and gain/dispersion bandwidth}

\begin{figure}[!h]
\centering
\includegraphics[width=0.90\textwidth]{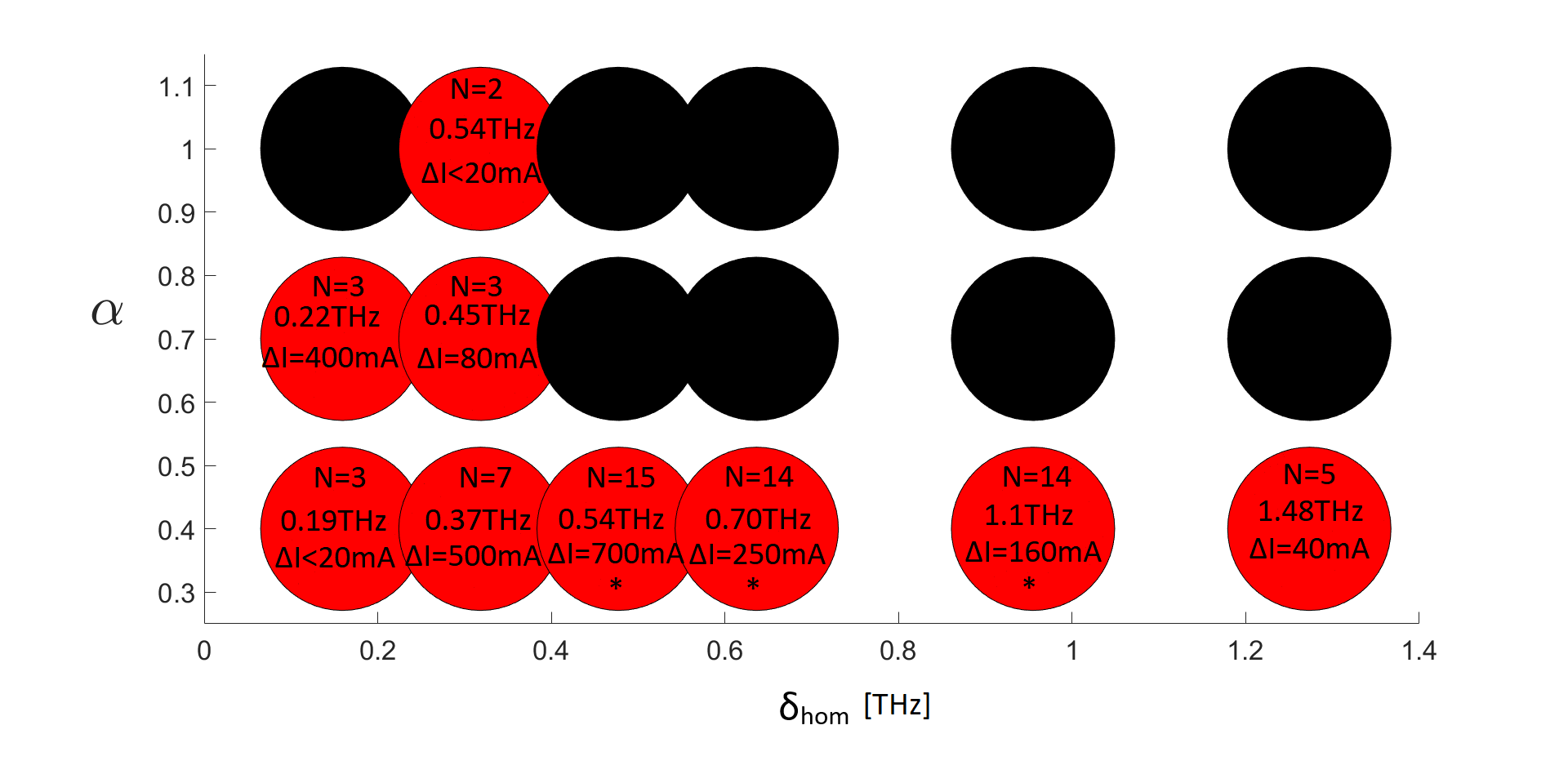}
\caption{Analysis of locked regimes upon variation of parameters $\delta_{hom}$ and $\alpha$. Black dots indicate that no locked regime could be found upon scanning the pump current in the interval $(I_{thr}, 3I_{thr}) $. Red dots indicate parameter pairs where such regime could be found. In the dots the dynamical FHWM gain linewidth (see text) in THz is reported along with the locking current range where locking was found $\Delta I$ and the corresponding value of $N_{10}$. The symbol '$*$' indicates the presence of more than one locking window.}
\label{GammaAlphaScan}
\end{figure}

In order to highlight the role of the LEF and the gain/refractive index dispersion  in affecting both the bias current range of the OFC regime and the figure of merit of  the optical comb, we run systematic sets of long ($>500ns$) simulations by sweeping the bias current between the threshold $I_{thr}$ and $3I_{thr}$ with a step of $0.19 I_{thr}$,  and considering $\alpha \in (0.4, 1)$ and $\delta_{hom}\in (0.16$THz, $1.27$THz$)$. The other parameters are those in Table 1.
Our results are conveniently summarized in Fig. \ref{GammaAlphaScan}, where we report for each pair $(\alpha, \delta_{hom})$ a black circle when no locking is observed, and a red circle in case of OFC emission; in the latter case inside the circle we also report the FWHM of the gain spectrum at threshold, the maximum number of locked modes found  in the $-10dB$ spectral bandwidth  and the extension  of the bias current interval $\Delta I$ where the OFC regime is found.


We first observe that spontaneous OFC formation is found diffusely throughout the considered values of $\alpha$ and of $\delta_{hom}$.
Also, as a general trend, in the locked regime the number of locked modes $N_{10}$ tends to increase with the FWHM of the gain curve.\\
We also report that, for a fixed value of $\delta_{hom}$, larger values of $\alpha$ increase the modal competition via nonlinear dispersion and reduce the range of $\Delta I$ where OFC is met in agreement with the results in \cite{columbo2018}. As an example, for e.g. $\delta_{hom}=0.32$THz where OFCs are reported for all values of $\alpha$, we found that $\Delta I$ drastically decreases as $\alpha$ increases. For fixed value of $\delta_{hom}$, the increase of LEF is equivalent to an increase of the asymmetry or inhomogeneity of the semiconductor material gain spectrum which is deviating from the ideal symmetric homogeneous gain of two-level atoms. On the contrary, low value of LEF implies a symmetric small inhomogeneous gain broadening, whereas the increase of $\delta_{hom}$ can be read as a reduction of the de-phasing time as typically observed increasing temperature.\\
At fixed $\alpha$, as a general trend an increment of $\delta_{hom}$ reduces the current range (or occurrence) for OFC regimes. These evidences seem consistent with the fact that the number of dispersed cavity modes for which the gain overcomes the losses increases with $\delta_{hom}$, but the quantity $N_{10}$ is actually limited by the efficiency of the FWM in locking the lasing modes that typically is an inverse function of distance from the resonance\cite{revcombs}.
In this regard an anomalous behaviour is found at the map edge where, for $\alpha=1$ and $\delta_{hom}=0.16$THz, we could not find any locked regime contrary to what happens for the two neighbouring circles of the map.
We may argue that this low value of the gain FWHM implies a destabilization of the single mode solution for high bias currents where the multimode regime is prone to be chaotic for the relatively high value $\alpha=1$. To corroborate this interpretation we checked that for $\alpha=1$ and $\delta_{hom}<0.16$THz only irregular multimode regimes are reliazed beyond the CW instability threshold.\\
Let us briefly analyze the results about the size $\Delta I$ of bias current generating the combs. If we focus on the case $\alpha=0.4$ where we report OFC formation for all the considered $\delta_{hom}$, for the lowest value of $\delta_{hom}$ we found a comb regime spanning just a few $mA$ in the whole simulation interval ($I_{thr}$, $3I_{thr}$); nevertheless, an extended comb regime of $\Delta I=1000mA$ can be found for higher values of the pump current ($I/I_{thr}>3$). For larger values of $\delta_{hom}$, $\Delta I$ keeps growing, it is maximum at $\delta_{hom}=0.48$THz and then decreases.\\

In order to  clarify the role of $\alpha$ in triggering the CW multimode  , we observe that it was already shown how increasing this parameter lowers the threshold for the multi-mode lasing (see Fig.3a in \cite{columbo2018}). In fact, since amplitude fluctuations lead to frequency fluctuations via $\alpha$, in presence of sufficiently large gain and bias current, we expect that a CW emission will be destabilized more easily in presence of larger $\alpha$. This mechanism is the only possible multimode  source in an unidirectional ring resonator, but in a FP configuration it would compete with SHB, a second well known mechanism for CW instability \cite{Gordon,Boiko}.

We numerically verified the previous considerations by simulating the QCL dynamics for $\alpha=0$ (ideal two-level system). We set $\delta_{hom}=0.48$THz, since it corresponds to the largest $\Delta I$ and maximum $N_{10}$ when $\alpha \neq 0$. In absence of SHB, we verified the expected CW emission even very far from threshold. We estimated the instability threshold (see chapter 20-22 in \cite{NOS}) and could verify that beyond that value ($I_{inst}>13I_{thr}$) a RNGH multimode instability sets in reducing our code to match the treatment of an unidirectional resonator and in the limit of small transmissivity. 
This result is consistent with the expectation that in  unidirectional, two-level case the well known RNGH instability is the only means to destabilize the single mode emission,  
triggered by the resonance of one cavity mode with the Rabi oscillation. 
By increasing  $\alpha$ (e.g. setting $\alpha=1.5$) and without SHB, we can confirm, in line with \cite{columbo2018,ArxivCapasso}, that the multimode instability affecting the single mode CW emission appears just above threshold. \\
When instead, keeping $\alpha=0$, and the SHB is switched on, we observe again CW destabilization just above the lasing threshold as we recently demonstrated for the QD laser case \cite{bardella2017}. We therefore conclude that either  the LEF or the SHB can (alone or together) contribute to the multi-mode emission which however does not necessarily lead to an OFC regime. The self-locked regime is found only for proper bias currents, for  proper combinations of LEF and homogeneous braodening linewidth and, as shown in the following, for fast enough carrier dynamics.  \\

\subsection{Pulses, chirping and OFC: the role of carrier dynamics}

\begin{figure}[h!tb]
\centering
\includegraphics[width=0.70\textwidth]{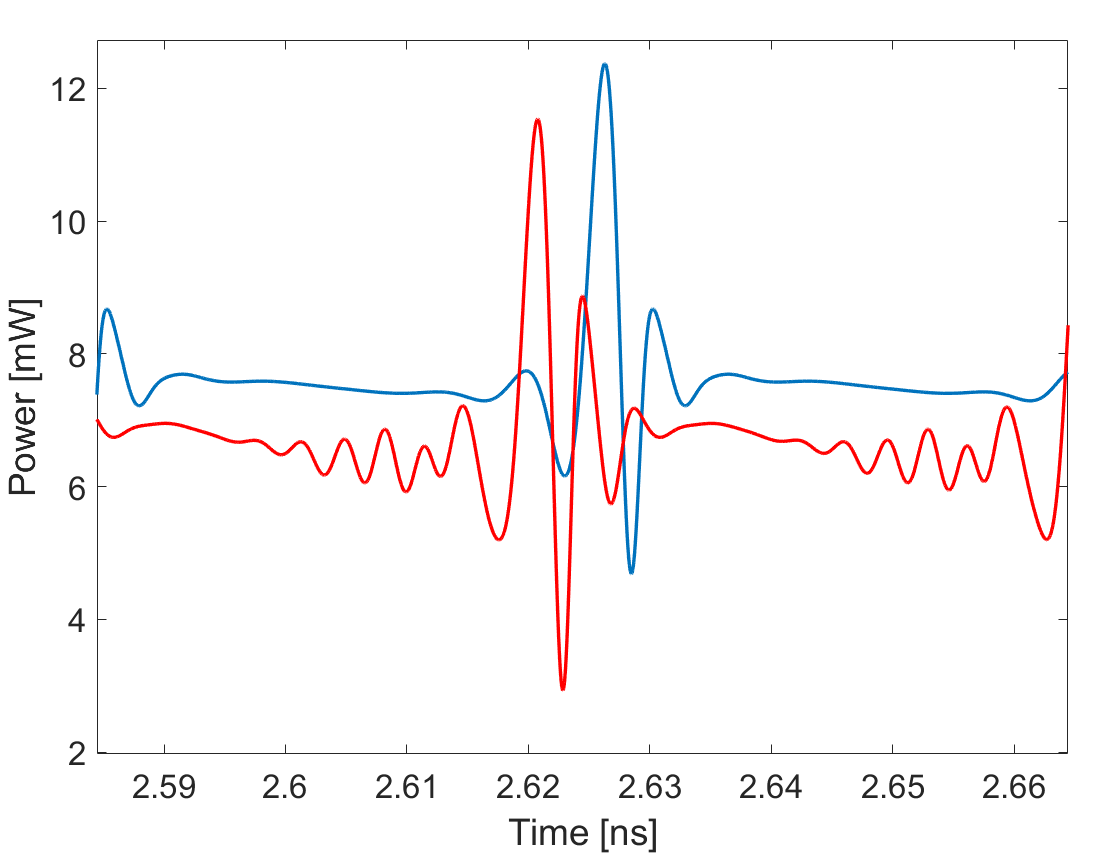} \caption{Zoom of a single power pulse for $\tau_e=1$ps (blue line) and $\tau_e=1.3$ps (red line) for $\delta_{hom}=0.48$THz, $\alpha=0.4$. The other parameters are as in Table 1. The width of blue pulse is estimated $25$ps, and $35$ps for the red one.}\label{impsovr}
\end{figure}

A relevant role in the formation of regular dynamics from multimode emission is played by the carrier decay time. In slow ($\tau_e \geqslant 100ps -1ns$) conventional semiconductor lasers (for example in quantum well laser diodes) the spontaneous OFC formation is scarcely reported. In agremement with that, our numerical simulations showed that increasing $\tau_{e}$ from $1ps$ to $1.3ps$ leads to a pulse broadening (Fig.\ref{impsovr}). For larger $\tau_{e}$, mode locking is lost for the same set of parameters of Fig.\ref{firstscan}.

\begin{figure}[!h]
\centering
\includegraphics[width=0.90\textwidth]{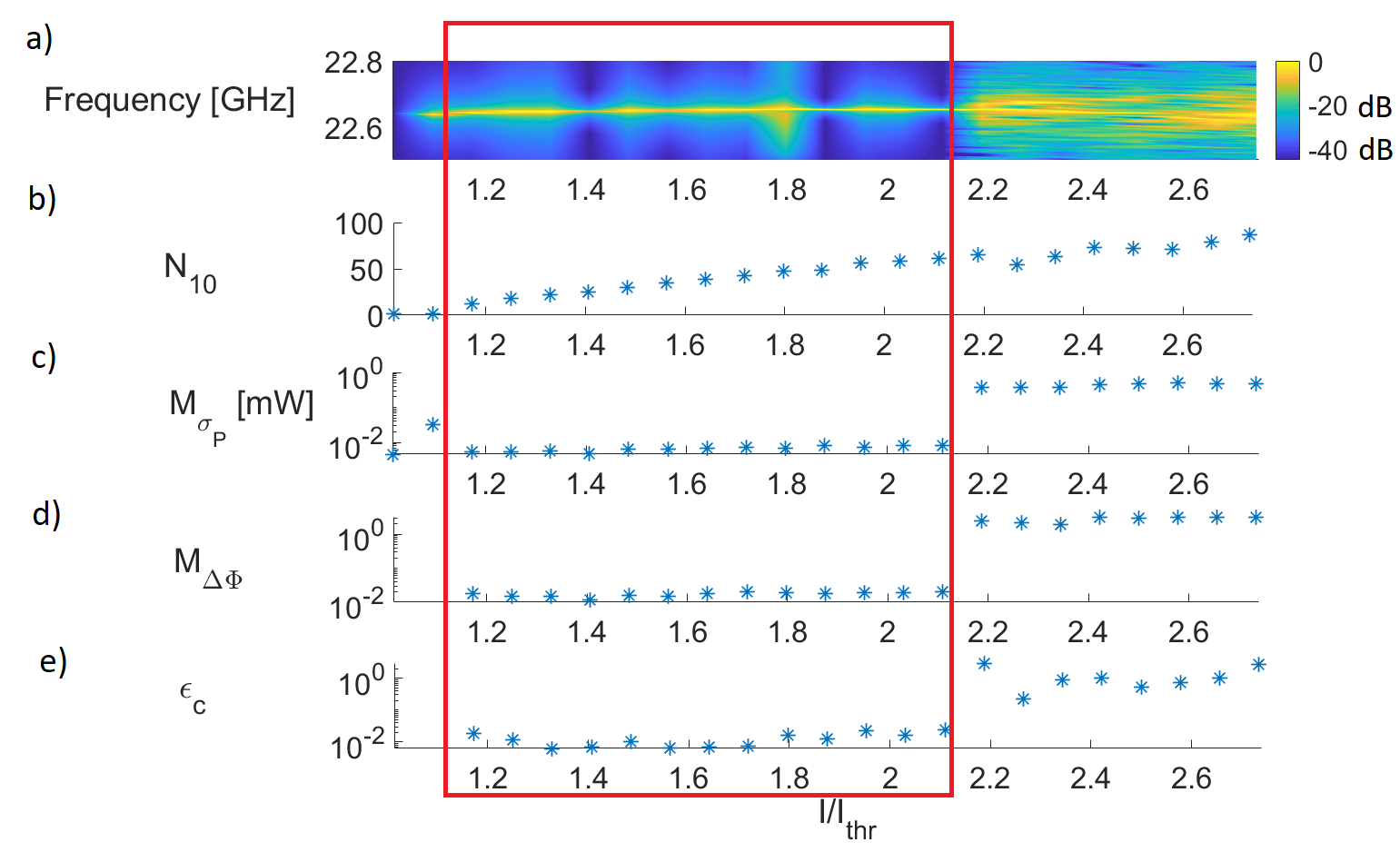}
\caption{a) Power spectrum map for the case $\tau_e=0.2ps$ with carrier grating. b) Number of modes in the -10dB band as a function of the current. c) $M_{\sigma_P}$ and d) $M_{\Delta\Phi}$ as functions of the ratio between bias current and threshold current. e) Chirp quantifier for the first $N_{c}=5$ Fourier coefficients of the instantaneous frequency signal.}
\label{taue02}
\end{figure}
\begin{figure}[!h]
\centering
\includegraphics[width=0.70\textwidth]{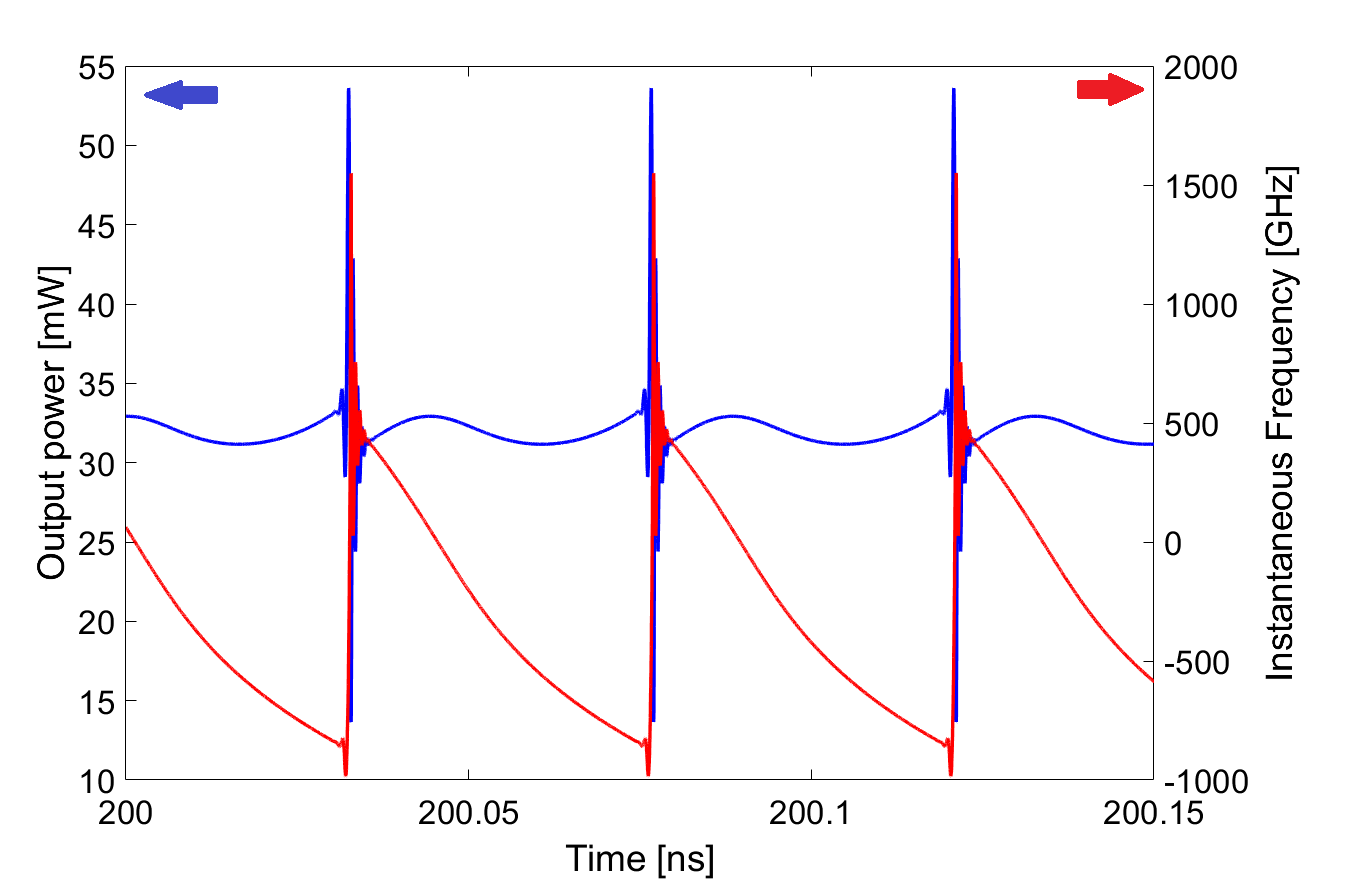}
\caption{Locked regime for $I/I_{thr}=2.16$, with $\alpha=0.4$, $\delta_{hom}=3.18THz$ and $\tau_e=0.2ps$. Temporal evolution of laser power (blue curve) and instantaneous frequency (red curve).}
\label{2700taue02}
\end{figure}

On the other direction, we  investigated the behaviour  for a fast carrier life time $\tau_{e}=0.2 ps$ (smaller than the value considered in previous sections). We also set  $\alpha=0.4$ and $\delta_{hom}=3.18$THz, which gives a FWHM of the gain bandwidth at threshold of $3.7 THz$, much larger then those considered in the map of Fig. \ref{GammaAlphaScan}. This gain bandwidth is comparable with the one measured in \cite{friedli}. We interestingly found that a reduction of the carrier lifetime is very beneficial in giving OFC regimes in quite wide bias current range and even for very large gain bandwidth FWHM.  Whereas the map of  Fig.\ref{GammaAlphaScan} shows that increasing the gain FWHM the OFC regime might be lost, we stress here that the OFC regime is also very dependent on the carrier lifetime. Thanks to the increased gain bandwidth we also observe a significant increase of the number of comb lines $N_{10}$. The OFC indicators versus bias current are in Fig.\ref{taue02}, where we see one very large comb region (red rectangle) characterized also by the presence of linear chirped regime, since $\epsilon_c<10^{-1}$ for all the current values in this region. The maximum number of locked modes  is $N_{10}=61$ found at $I/I_{thr}=2.16$; the corresponding  AM and FM dynamics at this bias current, shown in Fig.\ref{2700taue02}, shows shorter pulses and markedly linear chirps as compared to Fig.\ref{regular}.\\

\begin{figure}[!h]
\centering
\includegraphics[width=0.90\textwidth]{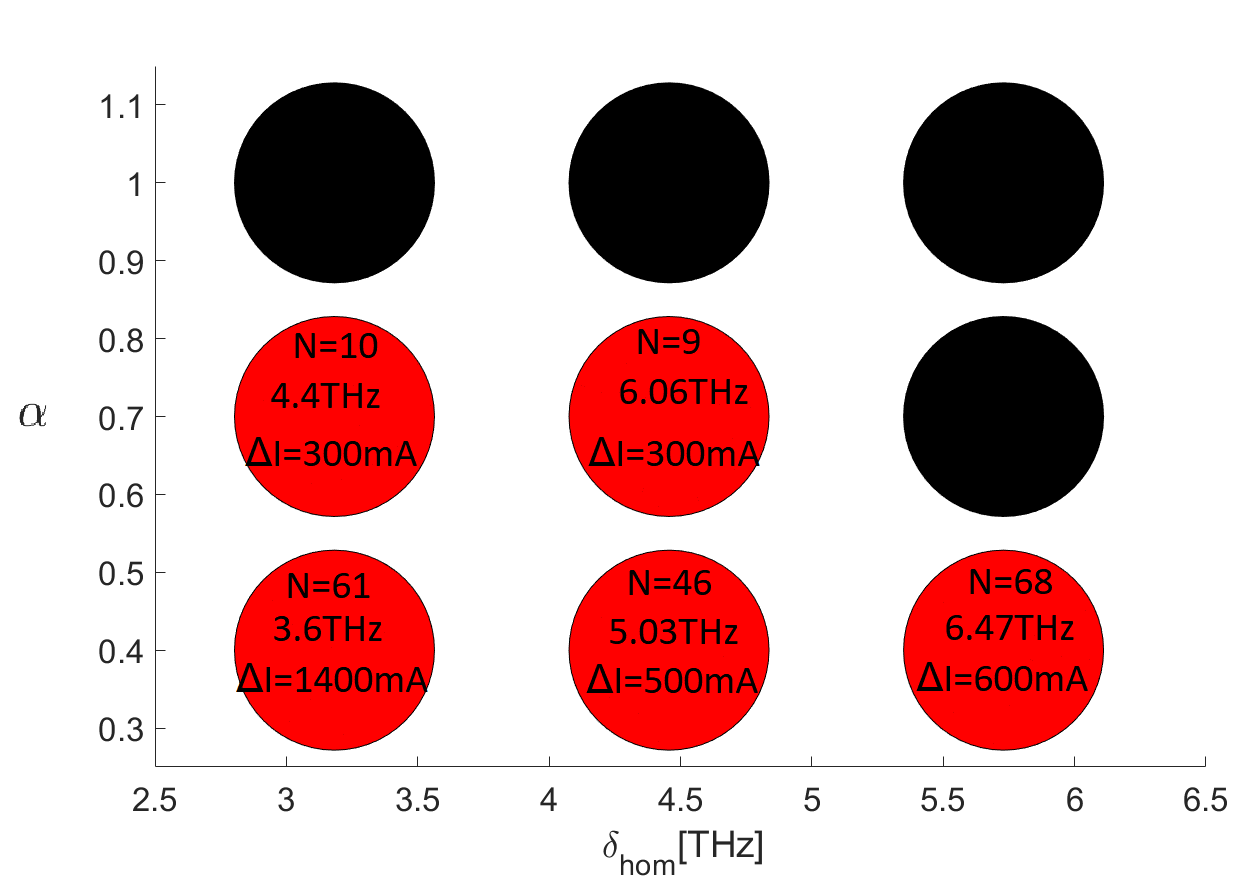}
\caption{Case $\tau_e=0.2ps$: analysis of locked regimes upon variation of parameters $\delta_{hom}$ and $\alpha$. Black dots indicate that no locked regime could be found upon scanning the pump current in the interval ($(I_{thr}, 3I_{thr} $). Red dots indicate parameter pairs where such regime could be found. In the dots the FWHM gain bandwidth (see text) in THz is reported along with the locking current range where locking was found and the corresponding value of $N_{10}$.}
\label{GammaAlphaScan02}
\end{figure}

Fig.\ref{GammaAlphaScan02} reports the map for $\tau_e=0.2ps$ in the parameter space $\alpha \in (0.4, 1)$ and $\delta_{hom}\in (3.18$THz, $5.74$THz$)$; for each parameter configuration the bias current has been scanned between $I_{thr}$ and $3I_{thr}$, with current step $0.08 I_{thr}$ of $100mA $. The other values are those in Table 1. For $\alpha=0.4$ we find locked cases for all the considered values of $\delta_{hom}$. The wider bias current range for OFC  corresponds to $\delta_{hom}=3.18$THz and the highest number of locked modes is achieved with a FWHM gain linedwith of $6.47$THz.
Locked states are found also for an higher (and probably more realistic) value of $\alpha=0.7$ \cite{aellen}, whereas locking is completely lost for $\alpha=1$. The trend is similar to the one in Fig.\ref{GammaAlphaScan}: the increase of the LEF causes a reduction of $N_{10}$ as well as a reduction of the bias current range of OFC operation.

\section{Conclusions}
In this paper we have presented results concerning spontaneous OFC obtained in an original model we developed to encompass critical features for the coherent multimode dynamics of a QCL such as 1) a FP resonator with counterpropagating fields, which allows to include SHB effect in the gain dynamics, 2) effective semiconductor medium dynamics which reproduce asymmetric gain and dispersion spectra.\\
Simulations correctly predict formation of OFC for bias currents close to lasing thresholds and, spanning the current up to a few time the threshold, they could also predict the recurrence of OFC ranges, spaced out by current intervals where modes delock and cause irregular field dynamics.\\
Our work is thus successful in providing a unique model capable of replicating the main evidences of several experiments in the field.\\
We have characterized the OFC regimes  and their dependence on the laser's gain bandwidth and LEF, finding in particular that an increase of the LEF, which corresponds to an increase of the phase-amplitude coupling, determines a reduction of the locking regime extension and the predominance of a chaotic behaviour and also implies a reduction of the number of locked modes. We qualified OFC regimes not only on the basis of a narrow BN spectral line (which is nevertheless a commonplace in experiments), but also observing reduced instantaneous frequency jitter and modal power fluctuations, as measured by purposely introduced quantifiers.\\
Another feature of our simulations is  the confirmation that OFC associated to a sufficiently large number of locked modes exhibit the propagation of well defined pulses inside the cavity (on an almost flat field background) and a linear chirping of the instantaneous frequency, which we also conveniently characterized. This allows us to evidence how AM and FM modulations of the emitted field are simultaneously present in OFC.\\
Finally, we investigated the role of carrier decay rates, i.e. the speed with which the medium evolves in time with respect to coherence and optical field, showing that faster carriers, with rates below $(1ps)^{-1}$, allow for shorter pulse formation in the OFC regimes, and, in association, for broader period of linear frequency chirping.\\
Having achieved such a powerful model opens a broad range of possible investigations aimed to improving the search for better-quality, more robust OFC existing in ever-wider current ranges. Also, we plan to extend our analyses towards devices where RF injection provides a forcing element for active frequency locking, as well as towards lasers with an external coherent injection, acting as an external control exploitable in principle for locking and structure formation addressing. 
On a more fundamental basis, the analysis of the instability leading to multimode emission in a QCL will be a focus of interest, since the characterization of phase/amplitude instabilities is crucial for the determination of the general dynamical behaviour of our optical system.




\end{document}